\documentclass[aps,prb,reprint,groupedaddress]{revtex4-1}
\usepackage{graphicx}
\usepackage{amsmath}
\usepackage{color}

\newcommand{\ddeg}{$^{\circ}$}

\begin{document}


\title{Interaction of chlorine with Stone-Wales defects in graphene and carbon nanotubes, and  {thermodynamical} prospects of chlorine-induced nanotube unzipping}


\author{M. Ij\"{a}s}
\email[]{mari.ijas@aalto.fi}
\author{P. Havu}
\author{A. Harju}
\affiliation{ COMP Centre of Excellence, Department of Applied Physics, Aalto University School of Science, FI-00076 Espoo, Finland}


\date{\today}

\begin{abstract}
We study the {binding of chlorine atoms to carbon nanotubes and graphene at a Stone-Wales defect}, and to the sidewalls of pristine nanotubes. We show using \emph{ab initio} thermodynamics that if the environment is chlorine-rich enough,  {the unzipping of carbon nanotubes into graphene nanoribbons with chlorinated edges may be thermodynamically feasible}. By controlling the Cl chemical potential through temperature and pressure,  {opening selectively tubes below a threshold diameter might be possible.}  {Additionally, we  find increased binding energies for chlorine atoms bound to Stone-Wales defects as compared} to defect-free graphene and nanotubes, indicating that defects might act as nucleation sites for tube unzipping. On graphene, no more than a single Cl atom may be bound to the defect in ambient conditions, limiting possible Cl-induced changes in the resulting nanoribbons. 
\end{abstract}


\maketitle


\section{Introduction}

Many of graphene's remarkable properties originate from the linear band dispersion forming the Dirac cones at the corners of the Brillouin zone. For some applications such as transistors, the opening of a band gap would be desirable. Two main methods have been suggested to open a gap in graphene. The use of lateral confinement into quasi-one-dimensional graphene nanoribbons (GNRs) or zero-dimensional graphene flakes was predicted~\cite{Nakada-Fujita} to introduce a gap long before graphene was prepared experimentally. Fine-tuning of the gap using confinement requires control of the edge structure with atomic precision, both regarding  {the lateral size of the finite graphene structure} and the edge termination. On the other hand, current top-down lithographic approaches have not yet reached this degree of detail, even though there are promising recent advances.\cite{Zhang-Yazyev} Recently developed bottom-up approaches\cite{Cai,Blankenburg-Cai,Ruffieux-Cai,Treier-Pignedoli} seem promising, as the  precursor molecule uniquely determines the edge structure of the nanographene fragment formed by on-surface polymerization. The method for the preparation of narrow armchair ribbons is well established~\cite{Cai,Blankenburg-Cai,Ruffieux-Cai} but, this far, a bottom-up synthesis for zigzag-edges is not available. 

Another way to change the electronic properties of graphene is chemical functionalization. In addition of opening a band gap, this method may also change the charge carrier concentration~\cite{Liu-Liu-Zhu}  and modify the mechanical properties of graphene.\cite{Topsakal} Also transport properties are modified, even in a way that allows one to identify the defect based on transmission through the graphene device.\cite{Saloriutta}  In addition to single atoms, more complicated chemical groups such as -COOH may be attached  and used as "bridges" for further functionalization.\cite{Banerjee-Hemraj-Benny} The main advantage of  the chemical route is its versatility.  Moreover, it is possible to selectively functionalize or defunctionalize graphene regions.\cite{Sessi,Li-Zhou} This approach could pave way to all-carbon circuitry through the combination of unfunctionalized, conducting graphene regions and functionalized, insulating or semiconducting regions seamlessly connected through the carbon backbone. 

Carbon nanotubes (CNTs) are the quasi-one-dimensional version of sp$^2$ carbon and they can be thought to be formed from GNRs by rolling the edges together. Recently, the opposite process, the formation of GNRs by tube unzipping, has been experimentally reported.\cite{Kosynkin} Due to the use of strong oxidants, the resulting ribbons contain rather many oxygen-rich groups, such as epoxides and carbonyls.\cite{Kosynkin}   Also hydrogen and fluorine have been suggested to unzip tubes,\cite{Wu-Tse-Jiang,Lu-Scudder-Kioussis,Tsetseris-Pantelides} but similar to the oxidative case, yield in experiments surface-functionalized ribbons.\cite{Talyzin,Valentini} It would thus be of interest for applications to find a reagent capable of  unzipping nanotubes that simultaneously could not bind to the basal plane of the resulting ribbons. 

The interaction of hydrogen and fluorine, both with graphene~\cite{Elias,Ryu,Nair,Robinson,Jeon-Lee,Zboril,Sofo,Leenaerts,Samarakoon-Chen,Sahin-Topsakal-Ciraci,Havu-Ijas-Harju,Klintenberg} and nanotubes,\cite{Khabashesku,Balasubramanian,Froudakis} has been extensively studied. The theoretical predictions on the opening of a band gap that depends on the functionalizing species and its coverage, have been verified in experiments on graphene.\cite{Elias,Zboril,Nair}  The next halogen, chlorine, has received significantly less attention. Recently, two different groups reported chlorine attachment to graphene using photochlorination~\cite{Li-Zhou} or chlorine plasma.\cite{Wu-Xie}  A few experiments report incorporation of chlorine at low content to partly oxidized tubes.\cite{Yu-Zhang-Wang,Brichka-Prikhodko} As there is a wealth of literature regarding CNT functionalization using other chemical species, we refer the reader to Refs.~\onlinecite{Khabashesku,Balasubramanian,Froudakis} for a review.

The theoretical considerations on the interaction between graphene and chlorine have been restricted to the binding of single Cl atoms to graphene~\cite{Wehling,Zhou-Zu, Ijas-Havu-Harju} or a CNT,\cite{Pan-Feng-Lin,Erbahar-Berber} a pair of Cl atoms on a CNT,\cite{Erbahar-Berber, Saha-Dinadayalane} or full or half Cl coverage of graphene.\cite{Klintenberg,Medeiros,Yang-Zhou} Very recently, also other low coverages were addressed.\cite{Sahin-Ciraci} The discrepancy between the theoretically predicted nonbonding lowest-energy structure\cite{Klintenberg,Wehling,Medeiros,Yang-Zhou} and the experimentally observed covalent attachment,\cite{Li-Zhou,Wu-Xie} associated with a change in the carbon hybridization from sp$^2$ to sp$^3$, was shown to originate from a structural instability of the non-bonding structure, as observed from imaginary phonon frequencies. The buckled chlorographene with covalent C-Cl bonds on both sides of the carbon plane was thus found to be the stable structure. {In most experiments, however, graphene lies on a substrate and chlorination from both sides is unlikely. The binding of Cl to graphene edges was recently studied,\cite{Ijas-Havu-Harju} and it was found to be favored over one-sided attachment to the basal plane, to the extent that the formation of Cl-containing edges was found to be  {thermodynamically} feasible in a wide range of chemical potentials of Cl.} In the case of nanotubes,  theoretical considerations  {of Cl binding to them} have been restricted to particular tube sizes,\cite{Pan-Feng-Lin,Erbahar-Berber} and the Cl binding properties as a function of the tube size have not been reported. 

The presence of topological defects in graphene-based materials is well-known, and has also been recently observed experimentally.\cite{Banhart,Meyer-Kisielowski} There is a large number of non-hexagonal rings at the grain boundaries and vacancies are known to stabilize by reorganization into a pentagon-nonagon pair.\cite{Banhart} The defect sites are more reactive~\cite{Banhart} due to the deviations from the optimal bond angle of the sp$^2$ carbon atoms, and possible dangling bonds at vacancy sites. The simplest topological defect is the Stone-Wales (SW) defect~\cite{Stone-Wales} that is formed by a 90\ddeg~rotation of a bond and does not require a change in the number of carbon atoms.  Its reactivity has been studied concentrating io the case of graphene hydrogenation,\cite{Boukhvalov-Katsnelson,Chen-Hu-Ouyang} and for CNT on gas molecules for sensing applications.\cite{Giannozzi-Car-Scoles,Picozzi-Santucci-Lozzi,Valentini-Mercuri-Armentano} In the case of CNTs, the multiple inequivalent orientations of the SW defect have in most cases been neglected, even though especially for small tubes the orientation directly affects the associated strain and thus also stability of the SW-bound adsorbant. To the best of our knowledge, the effect of lattice imperfections on Cl binding has not been considered neither for graphene nor for CNT, even though defected regions with attached chlorine were experimentally observed.\cite{Wu-Xie} 

In this paper, we address the prospect of using chlorine to unzip carbon nanotubes, and we show that under a chlorine-rich atmosphere, tube opening might be  {thermodynamically} feasible. As the Cl bound to planar graphene in ambient is unstable, chlorine atoms remain bound only at the edges of the unzipped ribbons. In addition, we study the chlorine binding properties of Stone-Wales defects both on graphene and in CNTs, and show that at most a single Cl atom remains in the planar configuration {after unzipping}. In addition, we report the Cl binding energies to pristine armchair and zigzag CNT as a function of the tube diameter. 

\section{Computational details}

All density-functional theory calculations were performed using the "FHI-aims" code developed at the Fritz-Haber Institute.\cite{AIMS} The Perdew-Burke-Ernzerhof (PBE) functional was used to describe exchange and correlation, and the Brillouin zone sampling was adjusted to the supercell size, corresponding to a 48$\times$48$\times$1 k-mesh in the graphene primitive cell. Total forces in the supercell were converged to less than 0.005~eV/\AA{}, and electronic degrees of freedom to $10^{-6}$~eV.  {All atoms in the supercell were allowed to relax. }Van der Waals interaction was taken into account using an approach by Tkatchenko and Scheffler.\cite{vdW} 

For the calculation of binding energies, both atomic (Cl) and molecular (Cl$_2$) chlorine were used as reference states. The binding energies were calculated as
\begin{equation}\label{eq:EB} E_{B, \mathrm{Cl}_x} = \frac{E_{\mathrm{Cl-gra}} - (E_{\mathrm{gra}}+\frac{N_{\mathrm{Cl}}}{x}E_{\mathrm{Cl}_x})}{N_{\mathrm{Cl}}}, \end{equation}
where $x \in \{1,2\}$, $E_{y}$ denotes the energy of  {species} $y$, and $N_{\mathrm{Cl}}$ is the number of chlorine atoms. $E_{\mathrm{Cl-gra}}$ denotes the total energy of the full simulational supercell, and $E_{gra}$ the corresponding supercell without chlorine. 

In \emph{ab initio} thermodynamics, the presence of a gaseous environment is taken into account through the corresponding chemical potentials,\cite{Wassmann-Seitsonen} which in turn depend on the partial pressure of the gas and temperature through the well-known formula $\mu_i(T,P_i) = \mu_i^{\circ}(T)+k_BT\ln{P_i/P^{\circ}}$. The Gibbs energy of formation $\Delta G$ is given by 
\begin{equation} \label{eq:Gibbs} \Delta G = E_{F, \mathrm{Cl}_x} -\frac{\rho_{\mathrm{H}}}{2}\mu_{\mathrm{H}_2}-\frac{\rho_{\mathrm{Cl}}}{x}\mu_{\mathrm{Cl}_x}, \end{equation}
where $E_{F, \mathrm{Cl}_x} = E_{B, \mathrm{Cl}_x}N_{\mathrm{Cl}}/L$,\cite{Ijas-Havu-Harju} $L$ is the length of the edge, and $\rho_y$ and $\mu_y$ are the edge density and the chemical potential of species $y$, respectively. The preferred structure at $(\mu_{\mathrm{H}_2}, \mu_{\mathrm{Cl}_x})$ is the one with the lowest $\Delta G$. 

\section{C\lowercase{l} binding to pristine nanotubes and the possibility of tube unzipping along the tube axis \label{sec:tubeCl}}

\begin{figure}
\includegraphics[width = 0.95\columnwidth]{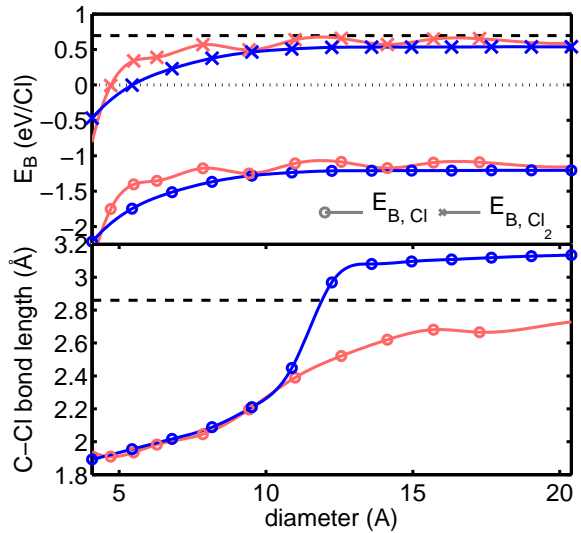}
\caption{\label{fig:zztube}(Color online) {Cl binding energy (upper panel) and the C-Cl bond length (lower panel) as a function of the tube diameter both for zigzag (red/gray) and armchair (blue/dark) nanotubes. In the upper panel, circles denote the binding energy with respect to atomic chlorine (Cl) and crosses with respect to molecular chlorine (Cl$_2$).  The dashed lines show the values from Ref.~\onlinecite{Ijas-Havu-Harju} corresponding to the 4$\times$4 graphene supercell, for which the Cl-Cl separation within the periodic boundary conditions is close to that in the nanotube supercell. The lines are to guide the eye. } }
\end{figure}

\begin{figure}
\includegraphics[width = 0.48\columnwidth]{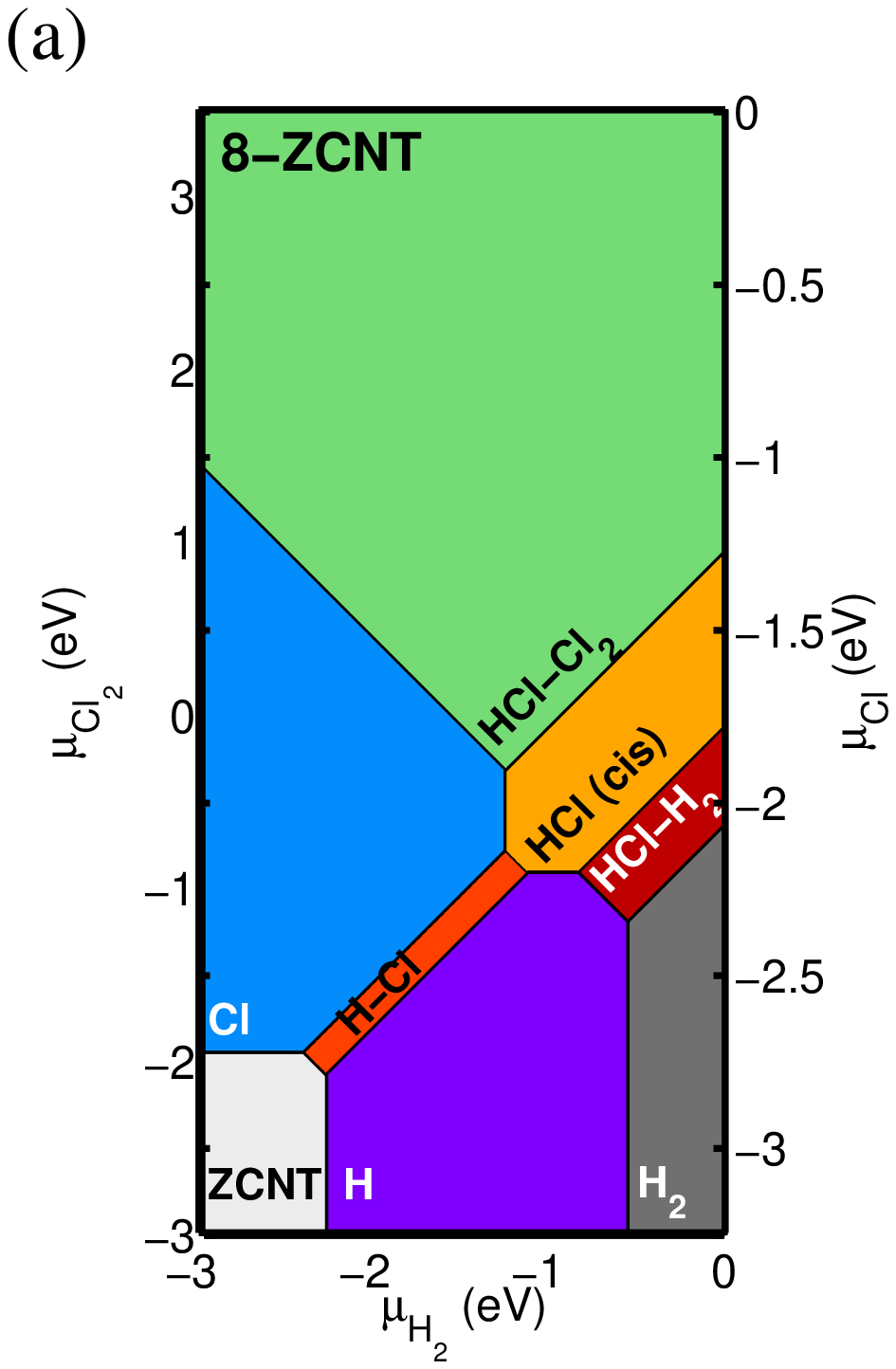}
\includegraphics[width = 0.48\columnwidth]{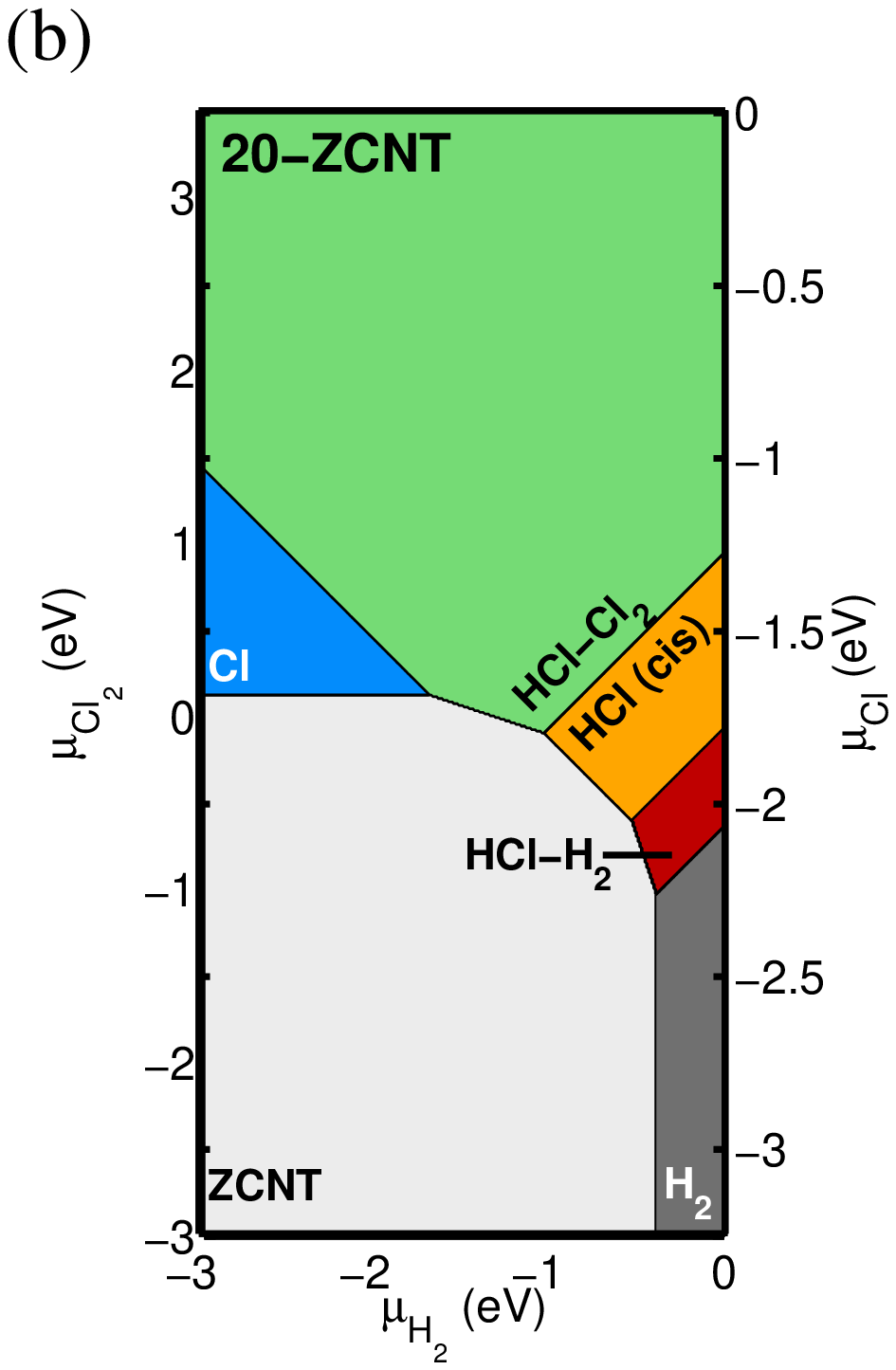}\\
\includegraphics[width = 0.48\columnwidth]{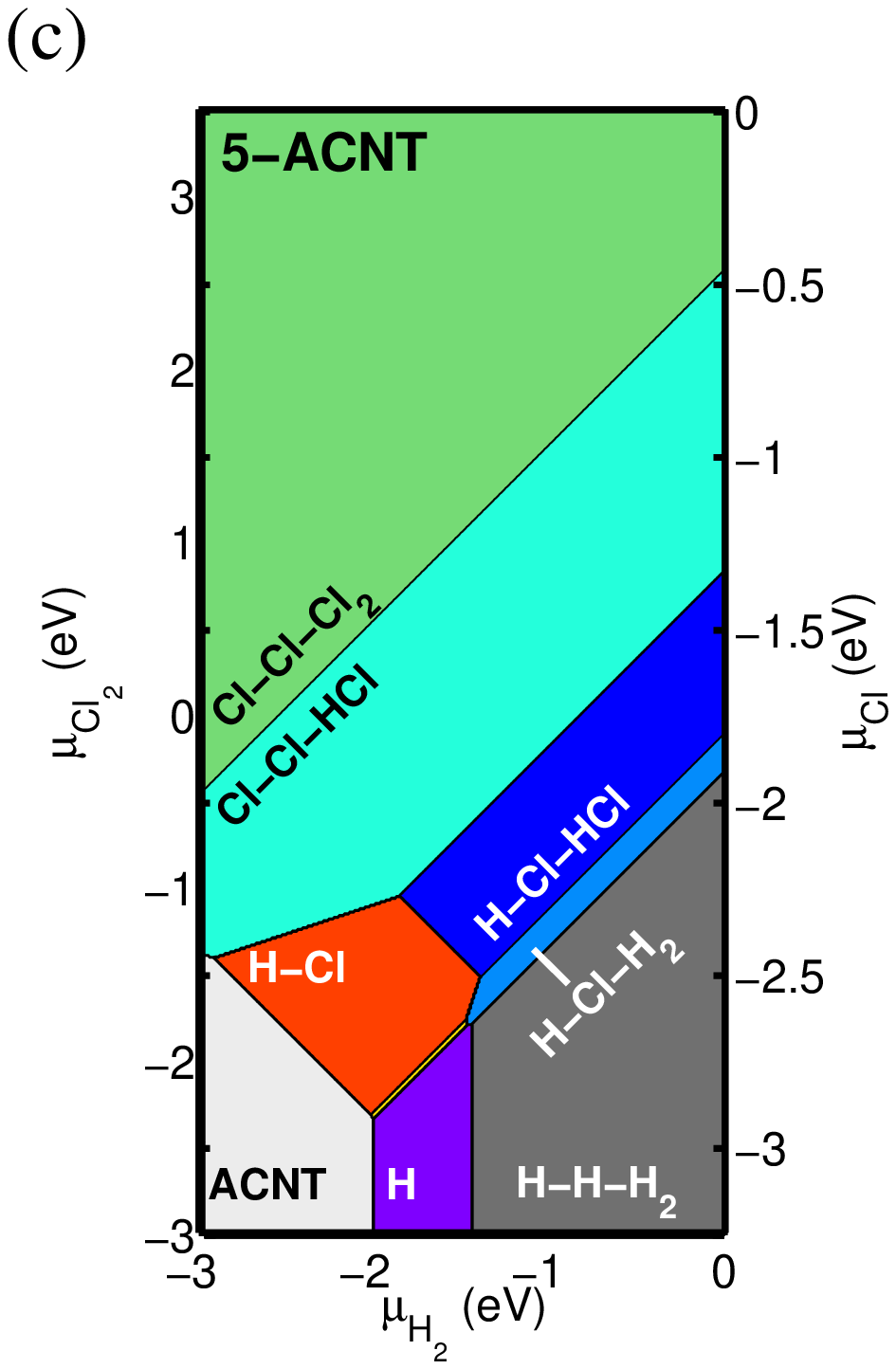}
\includegraphics[width = 0.48\columnwidth]{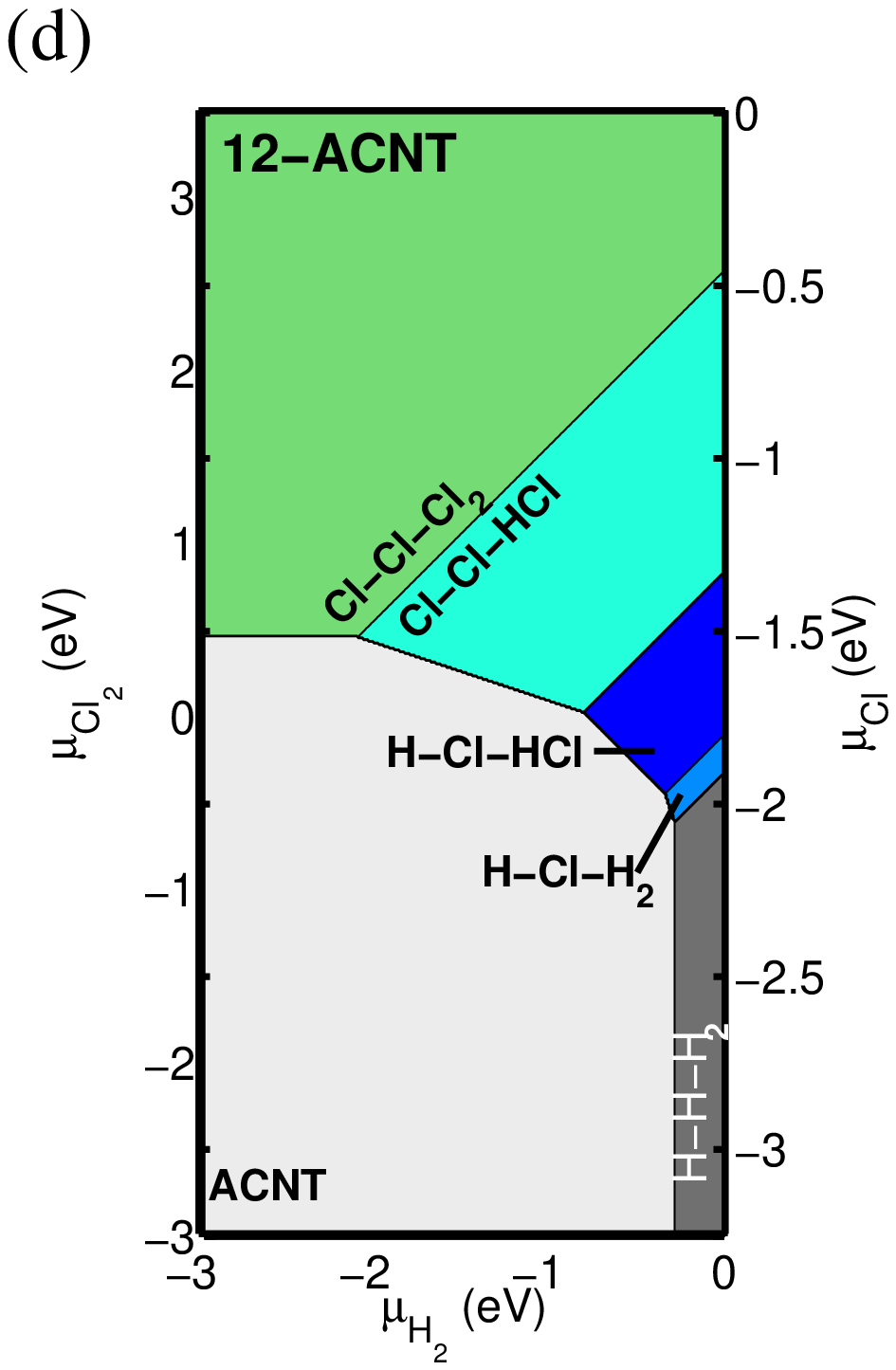}
\caption{\label{fig:CNT_stability} (Color online) The structures with the lowest Gibbs energy at different hydrogen and chlorine chemical potentials, $(\mu_{\mathrm{H}_2}, \mu_{\mathrm{Cl}_x})$. The naming convention of the different edge terminations is that of Ref.~\onlinecite{Ijas-Havu-Harju}.  (a) 8-ZCNT [$d\approx$6.3~\AA{}] (b) 20-ZCNT [$d\approx$15.6~\AA{}] (c) 5-ACNT [$d\approx$8.4~\AA{}] (d) 12-ACNT [$d\approx$16.3~\AA{}]}
\end{figure}

First, we address the binding of chlorine to pristine nanotubes. For simplicity and computational effort, we restrict our considerations to armchair and zigzag tubes  [($n,n$) and ($n,0$), respectively, using the usual CNT nomenclature].  Fig.~\ref{fig:zztube} (upper plot) shows the evolution of the Cl binding energy $E_B$ and the carbon-chlorine bond length  (lower plot) as a function of the tube diameter $d$, the distance between adjacent Cl atoms being approximately 8.5~\AA{} in ZCNT and 9.9~\AA{} in ACNT.  For reference, values for planar graphene from Ref.~\onlinecite{Ijas-Havu-Harju} are also shown for similar Cl-Cl separation. We see that with an increasing diameter, the binding energies approach the graphene limit, in the case of ZCNT superimposed with a periodic alternation due to the semiconducting and metallic families of zigzag nanotubes.  From Fig.~\ref{fig:zztube}(upper plot), we conclude that in atomic conditions, such as Cl plasma or the photochlorination experiment, binding is favored but in ambient, where Cl$_2$ formation is allowed, the attached Cl is unstable against desorption through Cl$_2$ formation, apart from the smallest tubes with  {diameter smaller than} $\approx 5$~\AA{}. 

The bond length shown in Fig.~\ref{fig:zztube} (lower plot) increases rapidly with tube diameter, and this increase can be associated with decreased covalent and increased ionic bonding character.  In the case of armchair tubes, there is a jump in the carbon-chlorine separation. This is due to the change of the preferred adsorption position from the top position in small-diameter tubes to a bridge position between two carbon atoms along the tube circumference in the larger tubes. 

Our results agree, in general, with literature for the tube sizes in which reference data are available. Pan~\emph{et al.}~\cite{Pan-Feng-Lin} obtained the bond length 2.09~\AA{} and binding energy -0.98~eV for a single Cl atom binding to 10-ZCNT [$d\approx$7.9~\AA{}]. Erbahar and Berber~\cite{Erbahar-Berber} found using LDA the C-Cl bond length of 1.93~\AA{} on 5-ACNT [$d\approx$8.4~\AA{}], and binding energy -1.73 -- -2.07~eV/Cl using atomic Cl as the reference state. Their binding energy is slightly lower than our, probably due to the overbinding tendency of LDA.  For the physisorption of a Cl$_2$ molecule, binding energy of -2.08~eV/Cl was reported~\cite{Erbahar-Berber} whereas we obtain -1.82 (-0.09) ~eV/Cl with respect to atomic (molecular) chlorine for Cl$_2$ on a ZCNT, and -1.83 (-0.09)~eV/Cl on a ACNT. It appears that the chirality of the tube has little effect on the adsorption energy of Cl$_2$. 

\begin{figure}
\includegraphics[width =0.95\columnwidth]{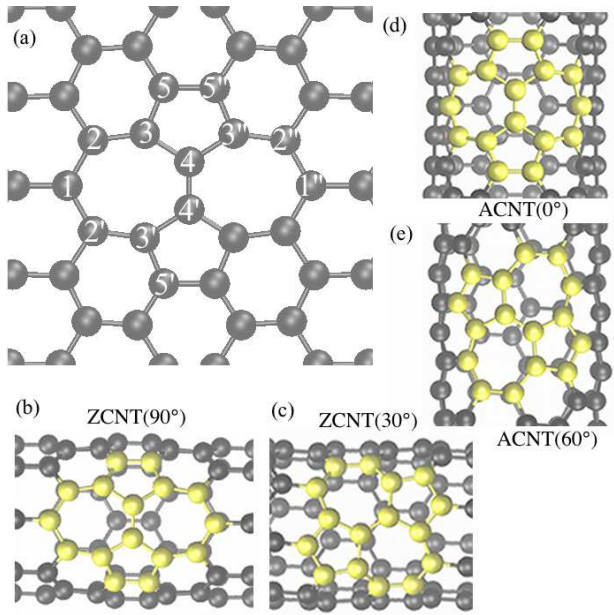}
\caption{\label{fig:SW}(Color online) (a) The different binding sites of a Stone-Wales defect. The numbers denote the inequivalent defect sites. '~and~'' give their symmetry-related counterparts.  (b) Zigzag CNT with Stone-Wales in 90\ddeg~orientation [ZCNT(90\ddeg)] (c) 30\ddeg~orientation [ZCNT(30\ddeg)] (d) Armchair CNT with Stone-Wales in 0\ddeg~orientation [ACNT(0\ddeg)] (e) 60\ddeg~orientation [ACNT(60\ddeg)]. For visual clarity, the carbon atoms belonging to the Stone-Wales defects have been colored yellow/light gray instead of black.   }
\end{figure}

\begin{table*}
\caption{\label{tab:SW_EB} The binding energies ($E_B$) and bond lengths ($r_{\mathrm{C-Cl}}$) for a single Cl atom bound to a Stone-Wales defect in graphene, and carbon nanotubes with different Stone-Wales orientations (see Fig.~\ref{fig:SW}) "-" means that the calculation started from the corresponding initial site leads to Cl diffusion along the carbon backbone onto another site.  The binding energies are given with respect to the Cl/Cl$_2$ reference state. The values for the most stable binding site for each system  {are} underlined. }
\begin{tabular}{c|cc|cc|cc|cc|cc}
 site &1& &2&&3&&4&&5&\\
&$E_B$ (eV)&$r_{\mathrm{C-Cl}}$ (\AA{})&$E_B$ (eV)&$r_{\mathrm{C-Cl}}$ (\AA{})&$E_B$ (eV)&$r_{\mathrm{C-Cl}}$ (\AA{})&$E_B$ (eV)&$r_{\mathrm{C-Cl}}$ (\AA{})&$E_B$ (eV)&$r_{\mathrm{C-Cl}}$ (\AA{})\\
\hline
graphene  &-1.46/0.28 &2.72 &- &- &-1.58/0.16 &2.14&\underline{-1.85/-0.11} &\underline{1.94} &-1.50/0.24 & 2.14  \\
ZCNT(90\ddeg)  &-1.68/0.06 &1.90 &-1.42/0.32 &1.92&-1.45/0.29 &1.94 &\underline{-2.08/-0.34} &	\underline{1.85} &-1.56/0.18 &1.92 \\
ZCNT(30\ddeg)  &-1.40/0.34& 2.02 &-1.46/0.29 &1.99&\underline{-1.99/-0.24} &\underline{1.86 }&- &- &-1.90/-0.16 &1.87 \\
ACNT(0\ddeg)  & -1.51/0.24 &2.27 &-1.73/0.01 &1.96&- &- &\underline{-1.98/-0.24} &\underline{1.91} &-1.75/-0.005 &1.97 \\
ACNT(60\ddeg)  &-1.65/0.09 &1.97 &-1.85/-0.11 &1.93&-1.74/0.002 &1.95 &\underline{-2.16/-0.42} &\underline{1.90} &-2.01/-0.27 &1.89 \\
\end{tabular}
\end{table*}

Recently, we addressed the  {thermodynamical} stability of graphene against the formation of chlorine-containing edges in different chlorine and hydrogen environments using \emph{ab initio} thermodynamics.\cite{Ijas-Havu-Harju} A similar approach may be used to address the possibility of unzipping CNT along their axis by using the energy of a carbon atom in a nanotube as the carbon reference instead of that in graphene. {Armchair and zigzag tubes {would} thus unzip into zigzag and armchair nanoribbons, respectively.} The nomenclature of Ref.~\onlinecite{Ijas-Havu-Harju} is used for the different Cl-containing edges.  {In the names of the edges, dashes separate adjacent edge carbon atoms. For instance, the notation "HCl-H$_2$" for an armchair edge means that H and Cl, and two H atoms are alternatingly attached to the armchair edge carbon atoms. "Cl-Cl-HCl" for a zigzag edge means that every third zigzag edge carbon binds both to H and Cl, and the two other only a single Cl atom. In HCl(cis), each armchair edge carbon binds both to H and Cl, the Cl atom being alternatingly above and below the carbon plane at adjacent edge carbons. } For illustrations on the chlorine-containing edge terminations, we refer the reader to Ref.~\onlinecite{Ijas-Havu-Harju}. 

From Eq.~(\ref{eq:Gibbs}), the Gibbs formation energy for the different edge terminations and different nanotube diameters can be calculated. Fig.~\ref{fig:CNT_stability} shows the stability diagrams in the ($\mu_{\mathrm{H}_2}$, $\mu_{\mathrm{Cl}_x}$) space for both armchair and zigzag CNT unzipping into zigzag and armchair tubes, respectively, at two different diameters. The more strained small-diameter tubes are clearly less stable and thus more prone to unzipping, as seen from the smaller regions in which the pristine CNT is the lowest-$\Delta G$ structure. Thus, by controlling the amount of Cl in the environment, it might be possible to selectively unzip only tubes smaller in diameter than a threshold value. Moreover, even for the larger-diameter tubes, at sufficient $\mu_{\mathrm{Cl}_x}$ there is a region in which chlorinated edges are  {thermodynamically} favored over the tubes, similar to the case of planar graphene. {As in ambient conditions Cl is weakly bound to the graphene basal plane and unstable against Cl$_2$ formation,\cite{Ijas-Havu-Harju} in unzipped nanoribbons, chlorine should be present only at the edges and thus not directly modify the graphene $\pi$ network.} 

Our approach neither considers the magnitude of reaction barriers nor gives information about the reaction path during the unzipping. The aim of the present study is to introduce the idea to be tested experimentally. As shown in Fig. 1, the binding of Cl to the tube is stronger at smaller tube diameters, changing the carbon hybridization from sp$^2$ toward sp$^3$, and elongating the C-C bonds close to the site of attachment. 
{From the results above, it is clear that  the presence of a single Cl atom is not enough to break the carbon-carbon bonds as it is bound or adsorbed on the CNT. } 
At finite temperature, thermal motion may well provide the energy needed to overcome the reaction barriers by opening these stretched bonds{, and additional chlorine atoms present may saturate the resulting dangling bonds.} In the case of open-ended tubes, it is also possible that Cl atoms diffuse to the inside of the tubes, and facilitate the binding by attaching to the sidewalls from inside. {In addition to sidewall-bound chlorine atoms, the opening could also nucleate at a defect.} In general, defects and imperfections are known to be more reactive than the pristine graphene lattice.\cite{Banhart} {The presence of the defect and Cl atom bound to it perturbs the CNT and might allow the binding of additional Cl atoms close by,  possibly initiating the unzipping process.} Thus, in the following, we address the interaction between chlorine and Stone-Wales defects, to see whether enough strain is induced to break some of the bonds, and to see if Cl bound to SW defects in CNT might remain on the unzipped ribbons.

\begin{table*}
\caption{\label{table:multi-Cl} The binding energies ($E_B$) and bond lengths ($d$) for a two Cl atoms bound to a Stone-Wales defect in graphene, and carbon nanotubes with different Stone-Wales orientations (see Fig.~\ref{fig:SW}). The three energetically most favorable configurations with chemical C-Cl binding are shown for each system, and the binding energy is given with respect to Cl/Cl$_2$. The site numbering is given in Fig.~\ref{fig:SW}. For graphene there are, in addition, configurations in which the Cl atoms combine into a Cl$_2$ molecule that physisorbs either on the defect, or on the pristine region nearby. }
\begin{tabular}{c|ccc|ccc|ccc}
			&sites & $E_B$ (eV)&$r_{\mathrm{C-Cl}}$ (\AA{})&sites & $E_B$ (eV)&$r_{\mathrm{C-Cl}}$ (\AA{})&sites &$E_B$ (eV)&$r_{\mathrm{C-Cl}}$ (\AA{}) \\
\hline
graphene 		 &3,3'' &-1.73/0.01 &1.92 &1,4 &-1.55/0.20 &1.92, 1.97 &5,4' &-1.47/0.27 &1.93, 1.95 \\
ZCNT(90\ddeg) &4,4' &-2.19/-0.44&1.82 & 5,4' &-2.01/-0.26 &1.84, 1.87 &1,4 &-1.92/-0.18 &1.86  \\
ZCNT(30\ddeg) &3,3'' &-2.17/-0.42&1.88 &4,4' &-2.09/-0.34 &1.86 & 5,4'&-1.85/-0.10 & 1.85, 1.96\\
ACNT(0\ddeg)   &3,3'' &-2.13/-0.39 &1.89 &3,4' &-2.09/-0.35 &1.88 &2,4' &-1.82/-0.08 &1.89, 2.00 \\
ACNT(60\ddeg) &5,4'  &-2.11/-0.37 &1.87, 1.88 &3,3' &-1.94/-0.20 &1.89, 1.90 &4,4' &-1.93/-0.19 &1.86 \\
\end{tabular}
\end{table*}

\section{C\lowercase{l} on Stone-Wales defects}

Again, we restrict our considerations to armchair and zigzag tubes as the two extremal cases  {for nanotube chiralities}, and use small tube diameters (5-ACNT: diameter 6.8~\AA{}  and 8-ZCNT: diameter 6.3~\AA{}). As seen in Sec.~\ref{sec:tubeCl}, the behavior of larger-diameter tubes approaches that of graphene, for which the computational supercell  of 8$\times$8 graphene primitive cells was chosen as a compromise between computational cost and accuracy. Previously, the chlorine adsorption energy on graphene was found to be fairly well converged for this supercell size.\cite{Ijas-Havu-Harju}  Five inequivalent binding sites on the defect were considered, as illustrated in Fig.~\ref{fig:SW}. As it has been established that halogens binding to graphene favor the "top" adsorption position,  both on graphene\cite{Ijas-Havu-Harju, Sahin-Ciraci} and on CNT,\cite{Erbahar-Berber} Cl atoms were initially placed on top of a carbon atom. As during the relaxation the Cl atoms diffused quite freely, in agreement with Ref.~\onlinecite{Sahin-Ciraci}, in which low diffusion barrier was found for chlorine atoms on planar graphene, bridge and hollow positions were not excluded. 

\begin{figure*}
\includegraphics[width = 1.99\columnwidth]{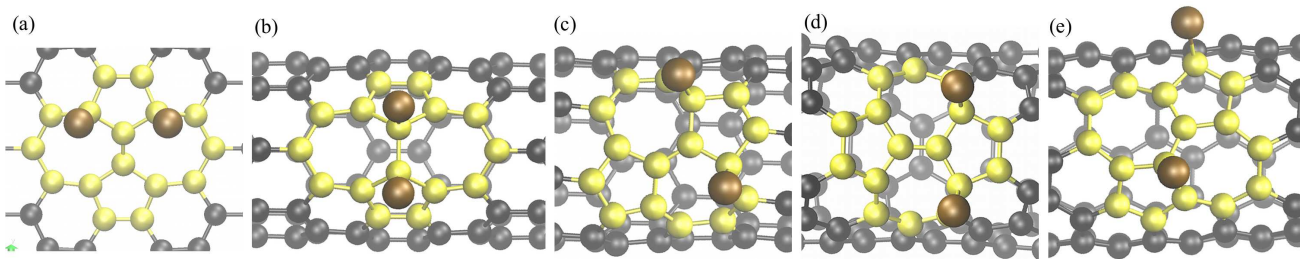}
\caption{\label{fig:SW_diCl} (Color online) The lowest-energy configurations for each SW system with two Cl atoms. (a) graphene (b) ZCNT(90\ddeg) (c) ZCNT(30\ddeg) (d) ACNT(0\ddeg) (e) ACNT(60\ddeg) For visual clarity, the carbon atoms belonging to the Stone-Wales defect have been colored yellow/light gray instead of black. Brown/gray larger spheres correspond to chlorine atoms. }
\end{figure*}

In nanotubes of general chirality, there are three inequivalent orientations of the carbon-carbon bonds, and thus three inequivalent orientations for the SW defects. In the achiral armchair and zigzag tubes, symmetry reduces these to two. The computational supercells were chosen such that the distance between the SW defects in ZCNT was 3 unit cells (12.8~\AA{}) and in ACNT six unit cells (14.8~\AA{}).  Fig.~\ref{fig:SW}(b)-\ref{fig:SW}(e) illustrate the two orientations with respect to the tube axis for both ACNT and ZCNT. We name the orientations using the approximate angle that the bond shared by the two pentagons forms with the tube axis, the orientations thus being ZCNT(90\ddeg), ZCNT(30\ddeg), ACNT(0\ddeg), and ACNT(60\ddeg). {In 8-ZCNT, for instance, the  ZCNT(90\ddeg) is 0.63 eV higher in energy than ZCNT(30\ddeg). For 5-ACNT, the orientation parallel to the axis [ACNT(0\ddeg)] is 0.83 eV higher in energy than the tilted ACNT(60\ddeg).} These differences in stability are easily understood as in ZCNT(90\ddeg) and ACNT(0\ddeg) orientations, the number of the more strained bonds lying approximately along the circumference of the tube is higher. The differences in the carbon-carbon bond lengths within the defect are within 0.05~\AA{} and depend on the tube chirality and SW orientation.

The results for the binding energies and C-Cl bond lengths of a single Cl atom on the defect are given in Table~\ref{tab:SW_EB}. In the case of tilted SW orientations, there are two inequivalent sites 2,3 and 5 in the structure. For instance in Fig.~\ref{fig:SW}(c), the bond between sites 3 and 4 can be oriented either along the tube circumference, or roughly along the tube axis. The results for the one with a lower $E_B$ are reported. In some calculations,  the chlorine atom diffused along the carbon backbone during the relaxation, and no binding energy for that initial site could be reported.

 In the case of graphene, we see that the binding to the defect site 4 is favored with $E_{B, \mathrm{Cl}}$/$E_{B, \mathrm{Cl}_2}$ = -1.85/-0.11~eV/Cl, in comparison to adsorption onto the pristine basal plane with $E_{B, \mathrm{Cl}}$/$E_{B, \mathrm{Cl}_2}$ = -1.25/0.50 eV/Cl  (12$\times$12 supercell).\cite{Ijas-Havu-Harju} Thus on pristine graphene, the binding of Cl is unstable in ambient conditions but a Cl atom bound to the SW defect is stable.  Site 4 has also been found to be the preferred binding site for a single hydrogen atom.\cite{Boukhvalov-Katsnelson,Chen-Hu-Ouyang}  Binding to the defect is clearly favored compared to the pristine sidewall also in CNT,  the binding energies for a pristine 8-ZCNT being $E_{B, \mathrm{Cl}}$ ($E_{B, \mathrm{Cl}_2}$) = -1.35 (0.39) eV/Cl and for a pristine 5-ACNT $E_{B, \mathrm{Cl}}$ ($E_{B, \mathrm{Cl}_2}$) = -1.51 (0.23) eV/Cl. 

In CNTs,  the orientation of the SW defect has a profound role on the energetic stability.  Even though binding to site 4 is, in general, quite energetic, for ZCNT(30\ddeg)  it is not the lowest-energy site. As easily seen in Fig.~\ref{fig:SW}, the preferred sites for each SW orientation apart from ACNT(0\ddeg) correspond to the sites that are part of a more strained bond nearly parallel to the tube circumference. Steric effects play a role for the stability, as depending on the orientation, the SW defect bulges slightly inwards or outwards. In ACNT(0\ddeg), chlorine atom actually pulls the bond between the two pentagons outward. 

We also consider the binding of multiple Cl atoms to a single SW defect, restricting the initial positions of the Cl atoms to those that belong to the defect, and including configurations in which two Cl atoms bind to SW sites that are equivalent in terms of symmetry distinguished by the use of primes (Fig.~\ref{fig:SW}). Fig.~\ref{fig:SW_diCl} shows the lowest-energy configurations for two chlorine atoms. On graphene, instead of covalent binding to graphene or CNT, Cl atoms frequently form Cl$_2$ molecules that are adsorbed onto to defect. In fact, adsorbed Cl$_2$ is more stable on and around the defect than on pristine graphene -- on graphene, the binding energy per Cl atom was found to be -0.125~eV/Cl,\cite{Ijas-Havu-Harju} whereas on the SW defect we find a value almost twice as large, $E_{B,\mathrm{Cl}_2}$ = -0.22 -- -0.25~eV/Cl depending on the exact position and orientation with respect to the defect. Interestingly, the lowest-energy configuration with two atoms does not contain the site preferred for the binding of a single Cl (site 4).  In some previous studies on hydrogen binding SW defect in graphene, it has been assumed that the lowest-energy single- {hydrogen} site is contained in the doubly  {hydrogenated} configuration.\cite{Boukhvalov-Katsnelson} 

One would expect that in CNT, chlorine would bind to both of the carbon atoms in the most strained bonds to provide maximal strain relief. This is indeed the case for outward-bulging ZCNT(90\ddeg), in which Cl atoms attach to the central rotated bond, whereas in the inward-bulging defects, this is sterically hindered. On other defected tubes, except ZCNT(90\ddeg), the separation between the two chlorine atoms is longer but still they tend to align rather along the circumference than along the tube axis but the separation between the chlorine atoms is longer. On the contrary to graphene, configurations with a Cl$_2$ molecule physisorbed onto the defect are more unstable than the covalently bound configurations. Table~\ref{table:multi-Cl} summarizes the three lowest-energy configurations with two Cl atoms for graphene and the four tube orientations.  

\begin{figure}
\begin{tabular}{cc}
\includegraphics[width = 0.47\columnwidth]{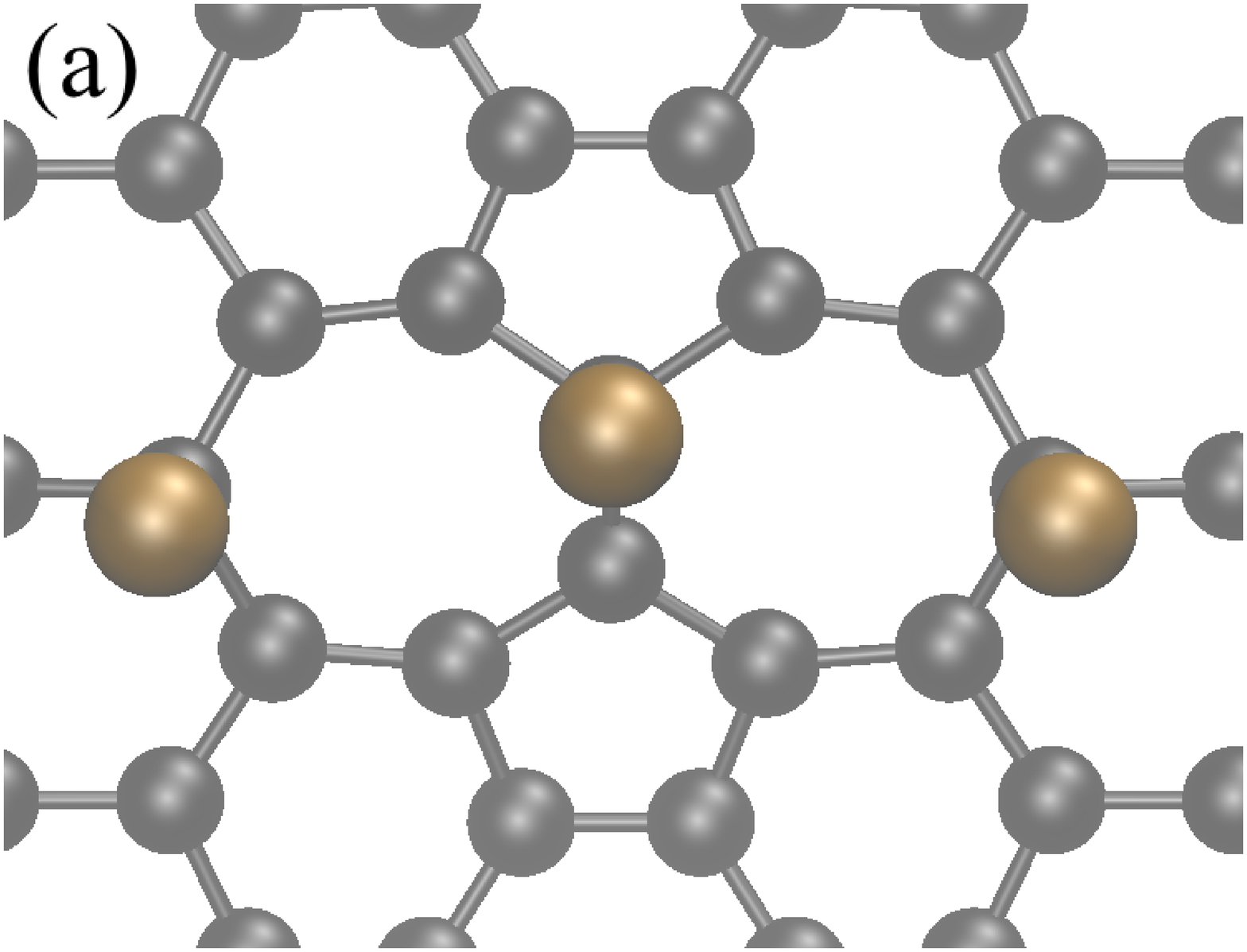}&
\includegraphics[width = 0.47\columnwidth]{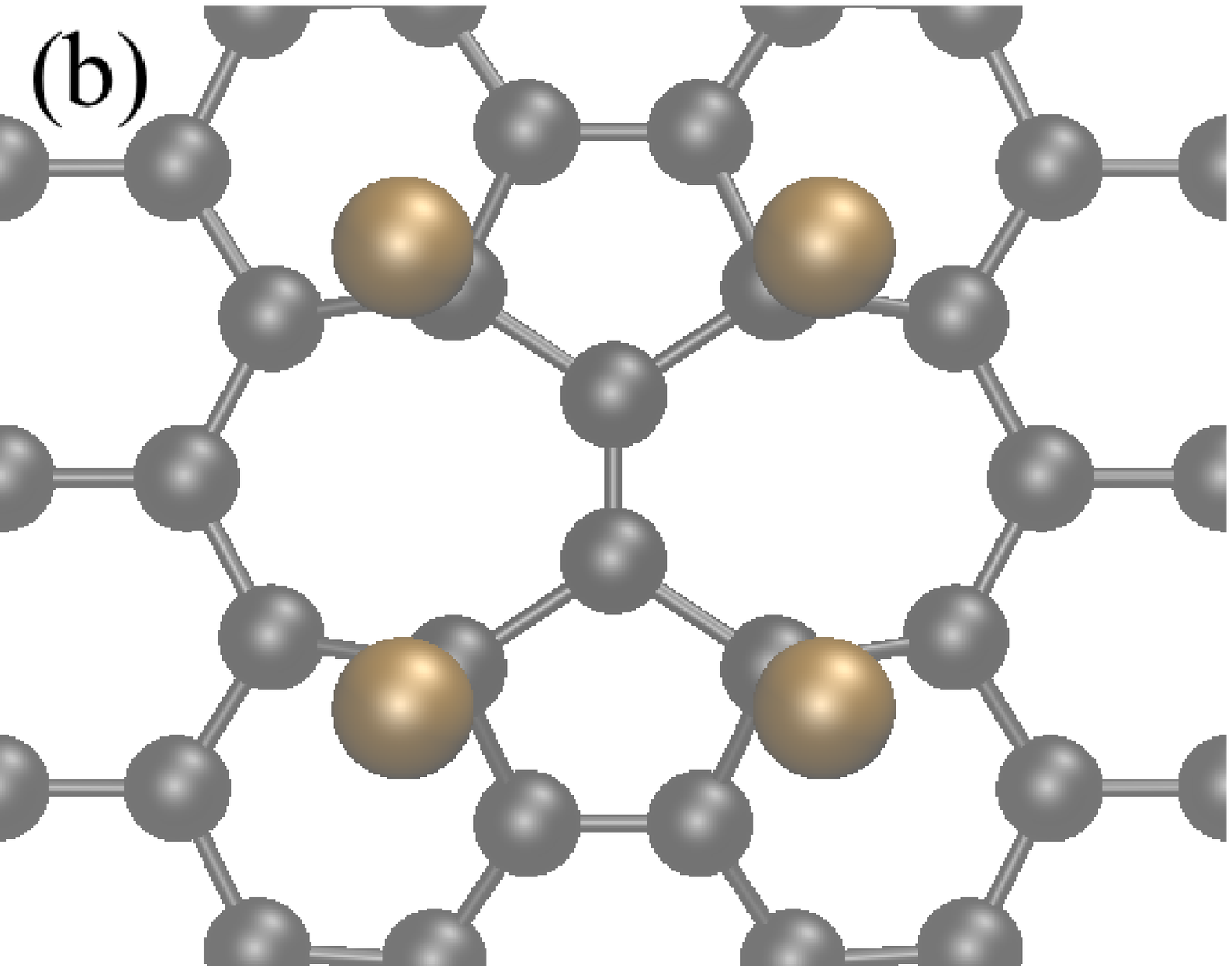}\\
$E_{B,Cl}$ = -1.40~eV/Cl &$E_{B,Cl}$ = -1.48~eV/Cl\\
$E_{B,Cl_2}$ = 0.34~eV/Cl &$E_{B,Cl_2}$ = 0.27~eV/Cl \\
$d_{C-Cl}$ = 2.01, 2.01, 1.89~\AA{} & $d_{C-Cl}$ = 1.93~\AA{}
\end{tabular}
\caption{\label{fig:graSW_multiCl} (Color online) The lowest-energy configuration and the binding energies $E_{B, \mathrm{Cl}_x}$ and C-Cl bond lengths $d$ for (a) three (b) four Cl atoms on the graphene SW defect.}
\end{figure}

Finally,  the binding of three or four Cl atoms on a single SW defect on planar graphene was considered. The lowest-energy configurations, along with their binding energies and bond lengths, are given in Fig.~\ref{fig:graSW_multiCl}. In these configurations, $E_B$ is clearly higher per Cl atom than in the doubly chlorinated defects. {Thus, dense clustering of chlorine atoms on defects is not feasible.}

\section{Conclusions}

We have studied the binding of Cl atoms to Stone-Wales defects in CNT and graphene, as well as to the pristine sidewalls of CNT as a function of tube diameter. We find that the binding of a Cl atom to the sidewall of the tube is energetically feasible in radical conditions with atomic chlorine but unstable in ambient that allows the formation of Cl$_2$ molecules, apart from the narrowest tubes with a diameter smaller than approximately 5~\AA{}. With increasing tube diameter, the binding energy and the bond length approaches {the ones found on} planar graphene. Using \emph{ab initio} thermodynamics, we show that in conditions with a sufficient Cl content in the environment, unzipping of the tubes into graphene nanoribbons with chlorinated edges might be  {thermodynamically} feasible, and in the planar geometry chlorine originally bound to the tube sidewalls is unstable and likely to desorb.  {We do not address the {unzipping} mechanism {of or the reaction barriers}, and introduce this idea for experimental testing.} 

{Based on the results on the binding of single Cl atoms onto the tube sidewalls, a single Cl atom binding to the sidewall seems unable to break C-C bonds, initiating the unzipping process. Defects perturb the hexagonal lattice, and might act as nucleation sites for unzipping. As an example, we} have also addressed the effect of Stone-Wales defects on the binding of chlorine. We find that chlorine prefers to attach onto the defects, and on nanotubes, attachment of two Cl atoms to a single defect is possible.  On planar graphene, Cl$_2$ is formed instead and thus after the unzipping,  the binding of only a single Cl atom on the defect is possible. 

\acknowledgements
M.I. acknowledges financial support from the V{\"{a}}is{\"{a}}l{\"{a}} foundation and from Finnish Doctoral Programme in Computational Sciences FICS. This research has also been supported by the Academy of Finland through its Centres of Excellence Program (project no. 251748). We acknowledge the computational resources provided by Aalto Science-IT project and Finland's IT Center for Science (CSC).

\bibliography{Cl_defects_final}

\end{document}